\documentclass[letterpaper,twocolumn,english,pre,showpacs,twocolumn,amsmath,amssymb,english]{revtex4}
\usepackage[utf8]{luainputenc}
\usepackage{color}
\usepackage{array}
\usepackage{graphicx}
\usepackage{csquotes}
\MakeOuterQuote{"}

\makeatletter

\pdfpageheight\paperheight
\pdfpagewidth\paperwidth

\@ifundefined{textcolor}{}
{%
 \definecolor{BLACK}{gray}{0}
 \definecolor{WHITE}{gray}{1}
 \definecolor{RED}{rgb}{1,0,0}
 \definecolor{GREEN}{rgb}{0,1,0}
 \definecolor{BLUE}{rgb}{0,0,1}
 \definecolor{CYAN}{cmyk}{1,0,0,0}
 \definecolor{MAGENTA}{cmyk}{0,1,0,0}
 \definecolor{YELLOW}{cmyk}{0,0,1,0}
}

\usepackage[nooneline]{subfigure}
\usepackage{dcolumn}
\usepackage{bm}
\usepackage{babel}

\usepackage{comment}

\makeatother

\begin{document}

\title{Enhancing Bremsstrahlung Production From Ultraintense Laser-Solid Interactions With Front Surface Structures}

\author{S. Jiang, A.G. Krygier, D.W. Schumacher, K.U. Akli, R.R. Freeman }

\affiliation{Physics Department, The Ohio State University, Columbus, OH, 43210,
USA}
\begin{abstract}
We report the results of a combined study of particle-in-cell and Monte Carlo modeling that investigates the production of Bremsstrahlung radiation produced when an ultraintense laser interacts with a tower-structured target. These targets are found to significantly narrow the electron angular distribution as well as produce significantly higher energies. These features combine to create a significant enhancement in directionality and energy of the Bremstrahlung radiation produced by a high-Z converter target. These studies employ short-pulse, high intensity laser pulses, and indicate that novel target design has potential to greatly enhance the yield and narrow the directionality of high energy electrons and $\gamma$-rays. We find that the peak $\gamma$-ray brightness for this source is 6.0$\times$10$^{19}$ ${\rm s^{-1}mm^{-2}mrad^{-2}}$ at 10MeV and 1.4$\times$10$^{19}$ ${\rm s^{-1}mm^{-2}mrad^{-2}}$ at 100MeV (0.1$\%$ bandwidth).
\end{abstract}


\maketitle

\section{Introduction}

Laser-produced X-ray and Bremsstrahlung sources have been a subject of significant research for several decades. This interest has been driven by a variety of applications including x-ray radiography of inertial confinement fusion plasma \textcolor{blue}{\cite{Perry:RSI1999}}, pair production \textcolor{blue}{\cite{Chen:PRL2009}}, and photonuclear physics \textcolor{blue}{\cite{Ledingham:PRL2000}}. These applications are typically driven by the hot electron beam which is produced when an intense laser interacts with a solid target. The hot electrons create incoherent radiation through several channels. First, the heated target produces blackbody radiation in the XUV range. Second, the hot electrons collide with background ions causing impact ionization and inner shell K line emission which is isotropic and generally monochromatic but broadens and shifts with heating \textcolor{blue}{\cite{Akli:POP2007}}. Finally, the hot electrons will suffer angular collisions with background ions which creates directional but broadband Bremsstrahlung radiation \textcolor{blue}{\cite{Edwards:APL2002,Courtois:POP2009,Murnane:POFB1991,Matthews:APL1983,Kmetec:PRL1992,LaFontaine:POP2012}}. In general, these processes can all be optimized by the ability to enhance the hot electron population. 

There have been numerous efforts to increase the number and energy of the hot electrons and also to increase the efficiency of the conversion of laser energy into hot electron energy. Most of these studies have emphasized the role of the laser pulse energy, duration, and intensity \textcolor{blue}{\cite{Wilks:1992,Beg:1997,Haines:2009,KlugeEnergy:2011,Krygier:POP2014}}; in addition, there is a robust literature describing the effect on electron energy of a "pre-plasma" on the front of the target \textcolor{blue}{\cite{Paradkar:2011,Scott:2012,Ovchinnikov:2013}}. 

There has also been a good deal of effort put into investigating the effect of surface roughness of targets on laser-plasma coupling \textcolor{blue}{\cite{Palchan:2007, Rajeev:2003, Sumeruk:2007, Kulcsar:2000,Murnane:1993,Kahaly:PRL2008,Hu:POP2010,Margarone:PRL2012}.} Some of these simulation works have found that the observed improvement is due to an effective surface area increase and the local field enhancement due to the roughness \textcolor{blue}{\cite{Kupersztych:POP2004, Klimo:NJP2011, Andreev:POP2011, Kemp:POP2013}}. Recently, several authors have shown that larger scale structures can give rise to enhanced production of hot electrons and/or high energy ions \textcolor{blue}{\cite{Kluge:2012,Gaillard:POP2011,Zigler:2013}}. Microcone targets were shot by Kluge and Gaillard who observed a significant increase in electron and proton energies which they attributed to direct laser acceleration (DLA) of electrons along the cone walls \textcolor{blue}{\cite{Kluge:2012,Gaillard:POP2011}}. A "slice cone" target, proposed by Zheng et al. \textcolor{blue}{\cite{Zheng:POP2011}}, showed similar results. Finally, similarly shaped nanobrush targets have been studied with 2D PIC modling \textcolor{blue}{\cite{Cao:POP2010_1,Cao:POP2010_2,Zhao:POP2010,Cao:POP2011,Wang:2012,Yu:POP2012,Yu:APL2012}}.

In previous work \textcolor{blue}{\cite{Jiang:PRE2014}}, we proposed a new method of enhancing the energy and directionality of the laser generated hot electrons. Rather than using higher energy, more intense lasers, we investigated novel target designs that are shown with 3D PIC simulations to dramatically alter and substantially increase the number of relativistic electrons produced in the laser-plasma interaction. In this work we expand on those results and investigate the production of Bremsstrahlung produced by using Monte Carlo simulations to propagate the PIC generated electron distribution through a high-Z converter. 

There are a variety of simple ways of manufacturing the structured targets which we are proposing including vapor-liquid-solid growth, metal-catalyzed chemical etching, molecular beam epitaxy, metal-organic chemical vapor deposition or deep reactive-ion etching \textcolor{blue}{\cite{Kayes:2007, Spurgeon:2008, Boettcher:2010, Kelzenberg:2010, Putnam:2010, Maiolo:2007, Peng:2005, Garnett:2008,Colombo:2009,Dong:2009,Goto:2009,Garnett:2010}}.  Unlike the microcone targets, the structures in our targets are periodic regular arrays; despite the difference we observe similar physics. The difference in the geometry has several benefits. First, experimental alignment of this target is similar to a standard flat foil and much easier than for the microcones. Second, guiding fields form that drive the electrons normal to the target base with narrow divergence, yielding a better radiation source. Finally, the results are somewhat robust against misalignment, unlike the microcone targets. We find that hot electrons are extracted from the tower structures and accelerated by the laser field; this is similar to that reported by some previous 2D simulations \textcolor{blue}{\cite{Kluge:2012, Gaillard:POP2011, Baton:POP2008, Psikal:POP2010, Micheau:POP2010}}. However, as we showed previously \textcolor{blue}{\cite{Jiang:PRE2014}}, fully 3D simulations are required to accurately treat the guiding fields.

We have performed a series of 3D PIC simulations that show the following features in the hot electron distribution. In every case, the targets with tower structures are found to produce a significantly greater number of higher energy electrons than for a standard flat target while maintaining the laser-to-electron conversion efficiency. The tower structured targets also show a significant narrowing of the electron angular distribution compared to flat targets, especially for high energies. Similar improvements are found when these electrons are used to generate high energy photons.

Experimental verification of these results should be relatively straight forward. First, production of this type of target has been demonstrated and features like the spacing and length of the tower structures are highly customizable. The electrons can only be directly measured after they have escaped the target, however, their angular distribution is heavily influenced by target charging and field distributions. Accordingly, verification should be done by measurement of high energy x-rays or gamma rays (using, for example, gamma-ray activation spectroscopy \textcolor{blue}{\cite{Sterlinski:1968}}) which correspond more closely with the laser-produced hot electron distribution than the escaped electrons. 
	
We describe the 3D PIC simulations in Section \textcolor{blue}{\ref{sec:sim_setup}} the results of which are given in Section \textcolor{blue}{\ref{sec:3Dresult}}. Section \textcolor{blue}{\ref{sec:xray}} describes the radiation production results and conclusions are given in Section \textcolor{blue}{\ref{sec:conclusion}}.

\section{Simulation Setup}
\label{sec:sim_setup}

\begin{figure}[pht]
\centering{}\includegraphics[width=75mm]{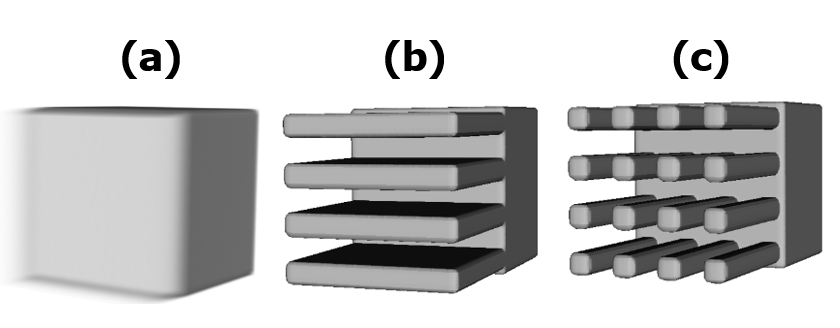} 
\caption{\label{fig:targets} (Color online) Sketch of the targets used in our simulations. In (\textcolor{blue}{a}) is a flat target with 1 $\mu$m pre-plasma. In (\textcolor{blue}{b}) is a target with slab structures on the front, the depth and width of the slabs are 10 $\mu$m and 1 $\mu$m. The spacing between the slabs is 2 $\mu$m. In (\textcolor{blue}{c}) is a target with tower structures on the front, the width of the towers is 1 $\mu$m. As shown in Table \textcolor{blue}{I}, the depth of the towers is usually 10 $\mu$m but is extended to 15 $\mu$m for one simulation. The spacing between towers in both transverse directions is 2 $\mu$m.}
\end{figure}

Our simulations use the 3D PIC code LSP \textcolor{blue}{\cite{Welch:2006}}. The targets we studied are shown in Fig. \textcolor{blue}{\ref{fig:targets}(a)} (flat type), \textcolor{blue}{(b)} (slab type) and \textcolor{blue}{(c)} (tower type).  Further, we have performed a moderate parameter scan on the tower targets.

\subsection{Parameters}

\label{sec:Our-simulations-use}

Fig. \textcolor{blue}{\ref{fig:targets}(a)} shows a normally flat target with an exponentially decaying pre-plasma on the front. The scale length of the pre-plasma is 1 $\mu$m. Fig. \textcolor{blue}{\ref{fig:targets}(b)} shows what we call a slab target, chosen for study as it is the 3D analog to the equivalent 2D tower target simulation. The slabs are 10 $\mu$m deep in the laser z direction, and 1 $\mu$m wide in the perpendicular laser polarization direction x. The spacing between the slabs is 2 $\mu$m. The third type is shown in Fig. \textcolor{blue}{\ref{fig:targets}(c)}, which we have labeled as a tower target. We have studied towers that are 10 $\mu$m or 15 $\mu$m deep and 1 $\mu$m wide. Here again the transverse spacings are 2 $\mu$m. Both structures in \textcolor{blue}{(b)} and \textcolor{blue}{(c)} have sharp interfaces. The base of the target for all but the low intensity case (IV) is 11 $\mu$m on a side; for the low intensity case, the base is increased to 23 $\mu$m to accommodate the larger focal spot. The material is Al, initialized as singly ionized, but further ionization is treated by the ADK ionization model; collisions are not included though we performed 2D simulations using Spitzer cross-sections with collision frequency capped to a maximum value of ${\rm 2}\times{\rm 10}^{16}$ ${\rm s}^{-1}$ and found no significant difference in either the spatial or angular hot electron distributions. The simulation box is made of ${\rm 120\times120\times600}$ cells with mesh sizes $\Delta$x=$\Delta$y=0.1 $\mu$m, $\Delta$z=0.05 $\mu$m. (We performed a resolution test using short simulations with ${\rm 120\times120\times1200}$ cells and $\Delta$z=0.025 $\mu$m. Neither the energies nor the trajectories showed significant differences compared to the coarser grid.) These simulations use absorbing boundaries and a 0.03 fs time step. 

We consider a Gaussian laser pulse that as part of our parameter scan is varied in peak intensity (by adjusting the beam waist). The constant properties of the laser are 15 J energy, 30 fs FWHM pulsewidth and 800 nm wavelength. For most simulations the intensity full-width at half maximum (FWHM) is 2.9 $\mu$m (corresponding to ${\rm 5\times10}^{21}$ ${\rm W/cm}^{2}$ peak intensity); one simulation uses a broader 9.1 $\mu$m intensity FWHM (corresponding to ${\rm 5\times 10}^{20}$ ${\rm W/cm}^{2}$ peak intensity). We also consider a variety of other conditions which are summarized in Table \textcolor{blue}{I}. Electron energy and spatial distributions are measured at a plane 5 $\mu$m inside the target. In all of our simulations reported here, we do not take into account changes to the electron spectrum due to target charging when the hot electrons leave the target, but we note that charging will have a minimal effect on the simulation for the higher energy (e.g. $>$ 100 MeV) hot electrons that are of interest here \textcolor{blue}{\cite{Link:POP2011}}. Finally, we employed a direct-implicit advance with an energy conserving particle push which greatly reduces numerical heating.

\begin{table}
\label{summary_table}

\begin{tabular}{c}
\includegraphics[width=75mm]{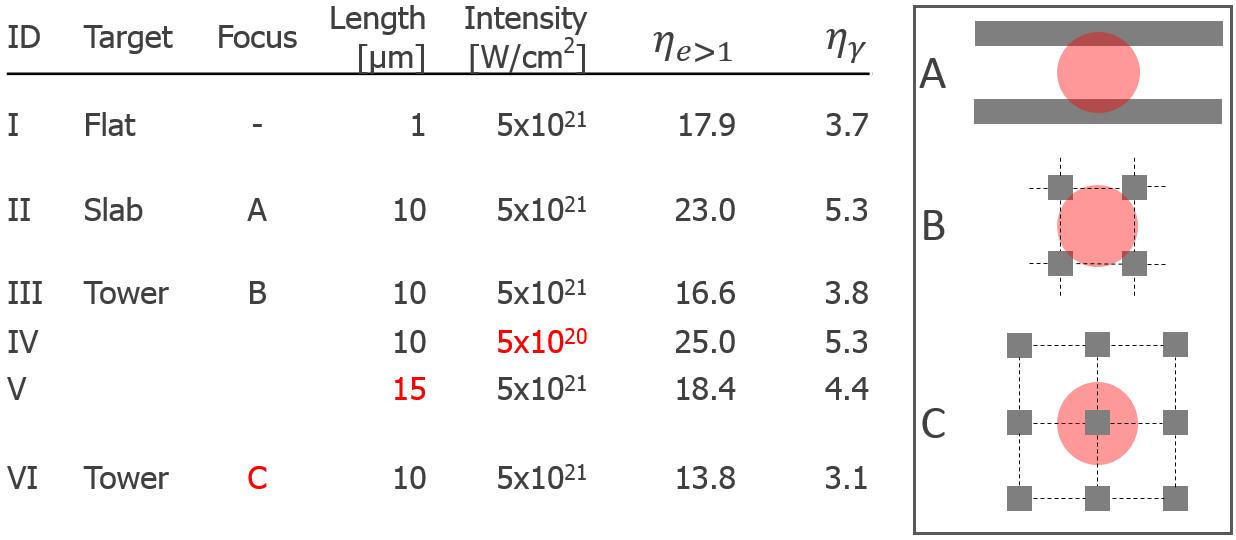}
\end{tabular}
\caption{Details of parameter scan and basic summary of results. The first column indicates the nomenclature for each case. The second column indicates the type of target studied. The third column gives the focal position of the laser relative to the structures - labels A, B, and C correspond to the sketches in the inset to the right. Focus C is used to examine the effect of experimental misalignment. The fourth column indicates the relevant length scale; for the flat this is the pre-plasma scale length and for the slab and towers it is the length of the feature. The peak laser intensity (set by the beam waist, holding all else constant) is given in the fifth column. The sixth column is the conversion efficiency from laser energy to electron energy, counting only those that cross a plane 5 $\mu$m into the target with kinetic energy larger than 1 MeV. The seventh column is the conversion efficiency from laser energy to $\gamma$-ray energy for a gold converter as discussed below in Section \textcolor{blue}{\ref{sec:xray}}.}
\end{table}

The choices for the parameter scan were made based on the results of a more extensive parameter scan using 2D PIC simulations. However, as we note in \textcolor{blue}{\cite{Jiang:PRE2014}}, 2D simulations are not capable of producing the results we find in 3D due to the inherent differences in the geometry. For instance, in a 2D simulation, the slab targets and tower targets have identical profiles while in 3D they produce significantly different results. Also, it is not possible to extract unambiguous conversion efficiencies in 2D. Furthermore, the angular spectra given below are found to have features which cannot be resolved in 2D (asymmetries out of the laser polarization plane). However, 3D simulations are much more computationally expensive so we have used 2D runs for preliminary study to select parameters for study in 3D.

Finally, we use the Monte Carlo particle transport code MCNP \textcolor{blue}{\cite{MCNPManual}} (which is also fully 3D but does not include fields of any kind) to propagate the laser-generated electrons through an Au converter target. All electrons in the PIC simulations that cross the plane 5 $\mu$m into the target (measured from surface of the "valley" between towers/slabs) are collected for propagation in MCNP. We then measure the angularly resolved energy distribution of the produced radiation. Due to the large fields produced in the interaction region and nearby inside the target, it would be preferred to generate the radiation consistently within the PIC simulations, however, using MCNP is significantly less costly. 

\section{3D PIC Simulation Results}
\label{sec:3Dresult}

\subsection{Energy Spectrum}

\begin{figure}[pht]
\centering{}\includegraphics[width=75mm]{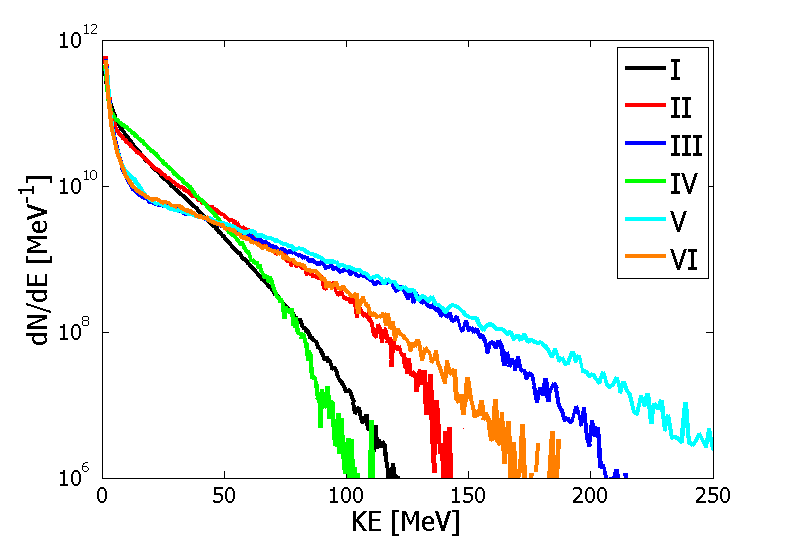} 
\caption{\label{fig:spectra} (Color online) The electron energy distribution for the targets used in our 3D PIC simulations. The naming convention of the legend is consistent with Table \textcolor{blue}{I}. In the high intensity cases (III, V and VI), there is a significant reshaping of the spectrum by the tower structures compared to the flat which produces a much larger number of electrons with energy larger than 50 MeV. The lower intensity case (IV) produces a similar result to the flat target but with higher overall conversion efficiency. }
\end{figure}

The electron energy spectra for the 3D simulations are shown in Fig. \textcolor{blue}{\ref{fig:spectra}}. The spectrum for the flat target (I) is given in black, the slab structured target (II) is in red, and the various tower structured targets are in blue, green, cyan and orange (III, IV, V and VI, respectively). We include a 1D exponential pre-plasma with a scale length of 1 $\mu$m in the flat target case to be more comparable to experimental situations while limiting the grid size; the other targets are sharp interfaces. Pre-plasma also increases the laser absorption and coupling to the target making the comparison more interesting. 

The conversion efficiency from laser energy to electrons with kinetic energy larger than 1 MeV is given in the sixth column of Table \textcolor{blue}{I}. The flat target produces a typical electron energy distribution with a coupling efficiency of 17.9$\%$. Compared to the flat, the slab target is found to produce a hotter spectrum as well as increase the overall coupling efficiency to 23.0$\%$. For the high laser intensity cases (III, V and VI), the spectrum generated by the tower targets is modified even further while maintaining a comparable conversion efficiency to the flat target. The low energy end of the tower target spectrum is significantly reduced compared to the flat target and \emph{there is a dramatically increased yield of high energy electrons}. The tower target up-converts many of the lower energy electrons compared to the flat creating a significant re-shaping of the energy spectrum while roughly maintaining overall conversion efficiency. The lower intensity case (IV) produces a similar energy distribution to the higher intensity flat case but, at 25$\%$ has a larger conversion efficiency which therefore relaxes the constraints on the laser.

\subsection{Electron Angular Distribution}

\label{angular distribution} 

\begin{figure*}[pht]
\centering{}\includegraphics[width=160mm]{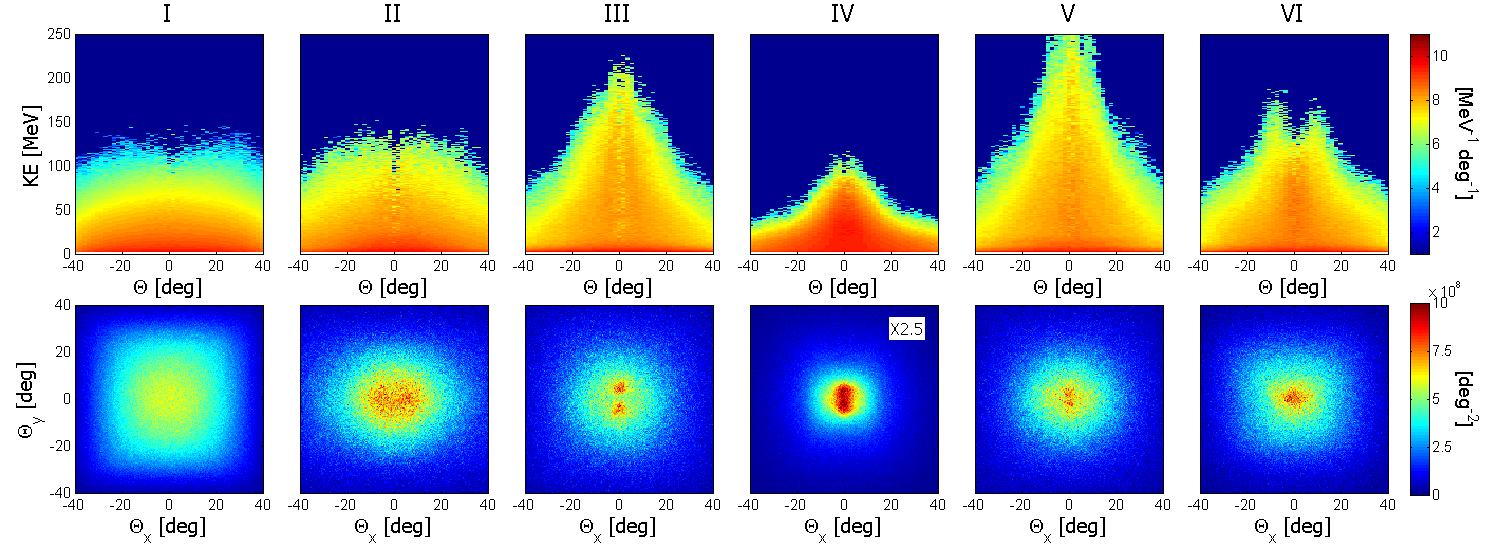} 
\caption{\label{fig:ang_distrib_3D} 
(Color online) Angular distribution of fast electrons ( $>$ 1 MeV). The top graphs are fast electron number distributions (on a color log scale) as a function of kinetic energy and angle $\theta$ with respect to the forward (z) direction. The bottom graphs are fast electron number distributions as a function of momentum direction with respect to the yz and xz planes. The titles of the top graphs indicate the simulation consistent with the naming convention in Table \textcolor{blue}{I}. For the flat target \textcolor{blue}{(I)}, an overall divergence angle of about ${\rm 60}^{\circ}$ is seen. The slab structured target \textcolor{blue}{(II)} is found to have an overall divergence angle of about ${\rm 40}^{\circ}$. The tower structured targets \textcolor{blue}{(III-VI)} show significant narrowing as well as some interesting features. In \textcolor{blue}{III}, the distribution shows two peaks at $\theta_{y}\approx\pm{\rm 4-5}^{\circ}$; for electrons $>$ 100 MeV, each cone angle is about ${\rm 4-5}^{\circ}$. In \textcolor{blue}{IV}, the low intensity case, energy distribution is cooler and the angular peaks disappear due to the reduced strength of the laser. In \textcolor{blue}{V}, the peaks are merged due to the longer length of the tower. In \textcolor{blue}{VI}, the misalignment modifies the angular distribution.} 
\end{figure*}

These targets also have some striking modifications to the angular distribution of the hot electrons, as shown in Fig. \textcolor{blue}{\ref{fig:ang_distrib_3D}}. The top row is electron number distribution as a function of kinetic energy and angle. The angle indicates the direction of electron velocities in a solid angle, $\Delta\Omega=2\pi \sin\theta\Delta\theta$, where $\theta=\tan^{-1}(\frac{\sqrt{p_{x}^{2}+p_{y}^{2}}}{p_{z}})$. The bottom row shows the divergence map (projected into 2D) of electrons with kinetic energy $>$ 1 MeV. These maps show the fast electron angular number distributions as a function of $\theta_{x}$ and $\theta_{y}$, where $\theta_{x}=\tan^{-1}(\frac{p_{x}}{p_{z}})$, $\theta_{y}=\tan^{-1}(\frac{p_{y}}{p_{z}})$. The titles of the top row of plots indicate the simulation parameters consistently with Table \textcolor{blue}{I}. 

Comparing cases \textcolor{blue}{I}, \textcolor{blue}{II} and \textcolor{blue}{III}, the electron divergence reduces with increasing electron energy. However, for the flat target, the decrease is small compared to the two types of structured targets. While some collimation is seen in the high energy portion of the electrons generated by the slab target (top of column \textcolor{blue}{II}), the tower target (top of column \textcolor{blue}{III}) shows a significant improvement in the collimation. This result is clearer in the bottom graphs of Fig. \textcolor{blue}{\ref{fig:ang_distrib_3D}}. The flat target \textcolor{blue}{(I)} shows a cylindrically symmetric angular distribution, with a large divergence angle of about ${\rm 60}^{\circ}$. Both the shape of the slab and tower structure breaks the rotational symmetry. For the slab target \textcolor{blue}{(II)} , the distribution is wider in the x direction than in the y direction. For the first tower target with focus type "B", although the target shape itself has ${\rm 90}^{\circ}$ rotational symmetry, the angular distribution does not. This is especially clear in \textcolor{blue}{III}, which features 2 distinct peaks centered at $\theta_{y}\approx\pm{\rm 4-5}^{\circ}$, each peak is about ${\rm 4-5}^{\circ}$ FWHM.  For the lower intensity case, in \textcolor{blue}{IV}, the electron energy distribution is cooler and the peaks are less distinct due to the lower laser intensity. In \textcolor{blue}{V}, the angular peaks have been merged due to the longer tower length; the peak energies are also higher for the same reason. In \textcolor{blue}{VI}, the laser misalignment has modified the angular distribution slightly as well as cooled the energy distribution. In all of the cases, the tower structures produce a significantly narrowed electron angular distribution compared to the flat target.

The angular distribution of the high intensity tower target \textcolor{blue}{(III)} has a remarkable feature, the highest energy electrons preferentially fall into two small cones.  Approximately 30$\%$ of the electrons (or ${\rm 5.7\times10}^{9}$ in number) with energy greater than 100 MeV fall into the two ${\rm 5}^{\circ}$ full angle cones. We find that for the tower target \textcolor{blue}{(III)} the fast electron bunch has a pulse duration of 42 fs; considering only electrons with energy greater than 100 MeV, the pulse is shortened to about 13 fs which gives an average current of 70 kA. Assuming a source diameter of approximately 3 $\mu$m, this electron source brightness at 100 MeV is on the order of ${\rm 10}^{23}$ ${\rm s^{-1}mm^{-2}mrad^{-2}}$ (0.1$\%$ bandwidth).  

The physics behind the enhancements by the tower targets is discussed thoroughly in \textcolor{blue}{\cite{Jiang:PRE2014}}. The enhancement to the energy distribution arises because electrons are injected directly into the laser field from the solid density target rather than from low density pre-plasma near the critical surface.  This enables larger energies than in the flat case for two reasons. First, the initially clean interface allows the laser to pull a large number of electrons out of the structures where they are accelerated to high energy by the laser through the valley regions until they reach the target and the laser is reflected. Not surprisingly, we find that the highest energy electrons originate at the ends of the towers; the end of the tower is the effective limit on the acceleration length, thus the tower ends are the origin of the largest energy electrons. Second, the gaps are initially free of particles thus minimizing the effect from increased phase velocity of light in plasma which enables the laser phase to significantly outpace the electron and reduces the maximum electron energies.

The narrowing of the electron angular distribution occurs due to guiding by the structures as explained in \textcolor{blue}{\cite{Jiang:PRE2014}}. The electrons are guided along the structures by counteracting quasistatic magnetic and electric fields which arise due to the large currents and charge separation in the plasma.

\section{X-Ray Production}
\label{sec:xray}

We now consider the potential of the structured targets for enhancing the production of Bremsstrahlung radiation. The tower targets are found to produce hotter electron energy spectra while maintaining the overall conversion efficiency. This enhancement will have a significant effect on Bremsstrahlung production which is sensitive to the electron energy but not on $K_{\alpha}$ emission which is far more sensitive to electron number than energy.  As a result, we only consider overall radiation production, which for a high-Z material is dominated by Bremsstrahlung.  To do this we use the Monte Carlo code MCNP to propagate the electrons generated in the above 3D PIC simulations through a high-Z (Au) converter.

\subsection{Thickness Optimization}

The optimal thickness for Bremsstrahlung production is determined by investigating the total energy (integrated over transverse planes) carried by radiation as a function of depth for a thick converter target. The dimensions of the converter target are 10 cm $\times$ 10 cm transverse $\times$ 1 cm longitudinal; each Monte Carlo study uses $10^{8}$ input particles (the total number of tracks is much larger due to secondary particles).  We find that in every case, the total energy into radiation is maximized between 2.7 mm and 4.2 mm; this is consistent with the well-known radiation length for Au of 3.3 mm. The peak conversion efficiencies for each case is given in the last column in Table \textcolor{blue}{I}; these peak values are determined mostly by the laser-to-electron conversion efficiency rather than by the electron spectral modifications. The high intensity slab target and the low intensity tower target cases have the highest conversion efficiency, about 40$\%$ higher than the high intensity flat target due to relatively larger surface areas covered by the laser focal spot. The other high intensity tower cases tend to fall in between, except for the misaligned one (case \textcolor{blue}{VI}), which has slightly smaller yield.

\subsection{Energy and Angular Distribution}

To calculate the energy and angular distributions of the radiation produced by this interaction, we have used a 1 cm $\times$ 1 cm transverse $\times$ 3 mm longitudinal converter target made of Au. The electron distributions generated in the PIC simulations are injected (preserving angular and energy information) into the front surface of the converter target; the angular and energy distributions are measured for escaping radiation. As before, a sampling of $10^{8}$ input electrons is used. Due to the limited spatial extent of the PIC simulations, the number of electrons with velocity polar angle larger than $50^{\circ}$ is slightly underestimated in this sampling; we expect this to have minimal effect on the results due to the directional nature of the electron source.

\begin{figure}[pht]
\centering{}\includegraphics[width=75mm]{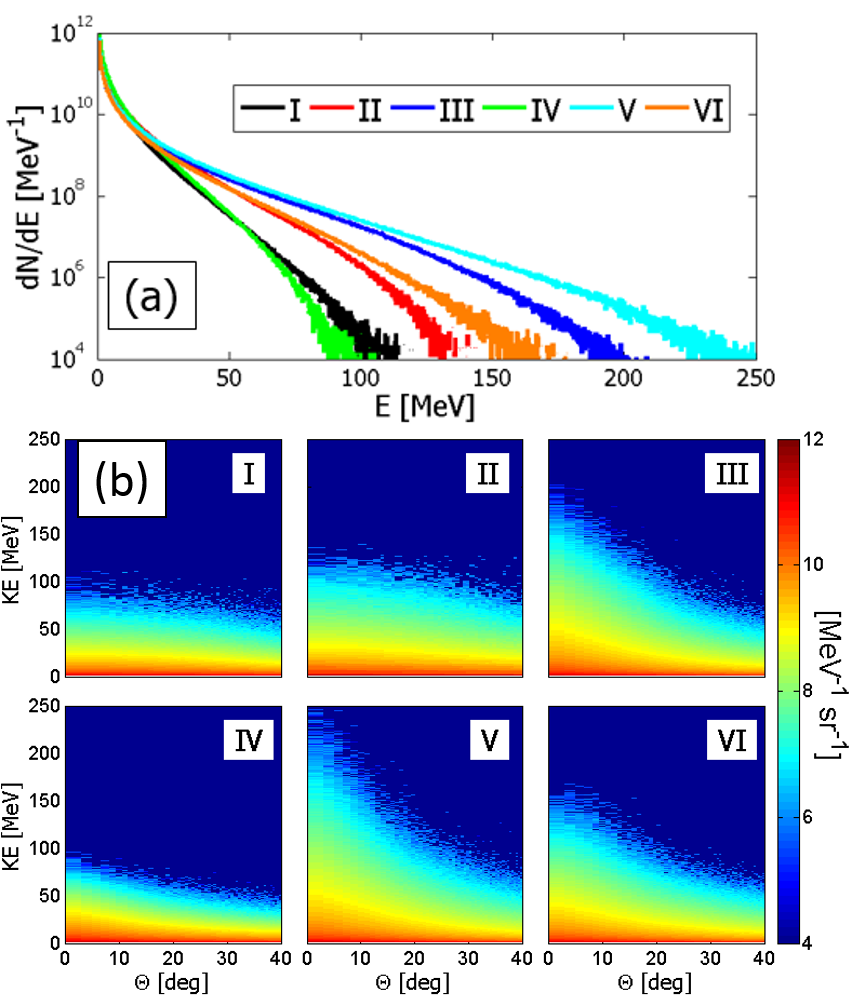} 
\caption{\label{fig:Xray_ang_spec} (Color online) The energy \textcolor{blue}{(a)} and angular \textcolor{blue}{(b)} distributions for the radiation generated using Monte Carlo simulations with PIC simulation generated electrons for the conditions described in Table \textcolor{blue}{I}. As expected, the trends are similar to the results for the electron distributions. The high intensity tower structures \textcolor{blue}{III} and \textcolor{blue}{V} show the highest energies. The cutoff energy from \textcolor{blue}{V} is higher than \textcolor{blue}{III} due to longer tower length. Compared to \textcolor{blue}{III}, the misaligned tower \textcolor{blue}{VI} generates lower energy X-rays, but they are still higher than other target types. The low intensity tower structure shows a similar energy distribution as the high intensity flat target, but the angular distribution is narrowed by a factor of 2 at almost all of the energy levels. Other tower targets at high intensities have slightly larger angles than the low intensity tower in its energy range, but much more collimated than the flat. At higher energies, the high intensity towers have further narrower distributions, the FWHM divergence angle is around or smaller than 10$^{\circ}$.}
\end{figure}

The radiation energy and angular distributions are shown in Fig. \textcolor{blue}{\ref{fig:Xray_ang_spec} (a)} and \textcolor{blue}{(b)}. Similar to the electron spectra, the towers with high intensities have the highest Bremsstrahlung energy. The 15 $\mu$m long tower produces a higher peak energy compared to the 10 $\mu$m long tower but the enhancement in the yield of X-rays $<$ 150MeV is not significant which indicates near optimization. The slab target, although producing the greatest total X-ray yield, generates lower energy radiation compared to the tower targets at the same intensity, even if the tower target is misaligned as in case \textcolor{blue}{VI}. The tower target at low intensity \textcolor{blue}{(IV)} generates a similar energy spectrum as the flat target at high intensity, but the angular distribution is much narrower - by about a factor of 2 in all energy ranges.

\begin{table}[pht]
\label{bright_table}

\begin{tabular}{c}
\includegraphics[width=75mm]{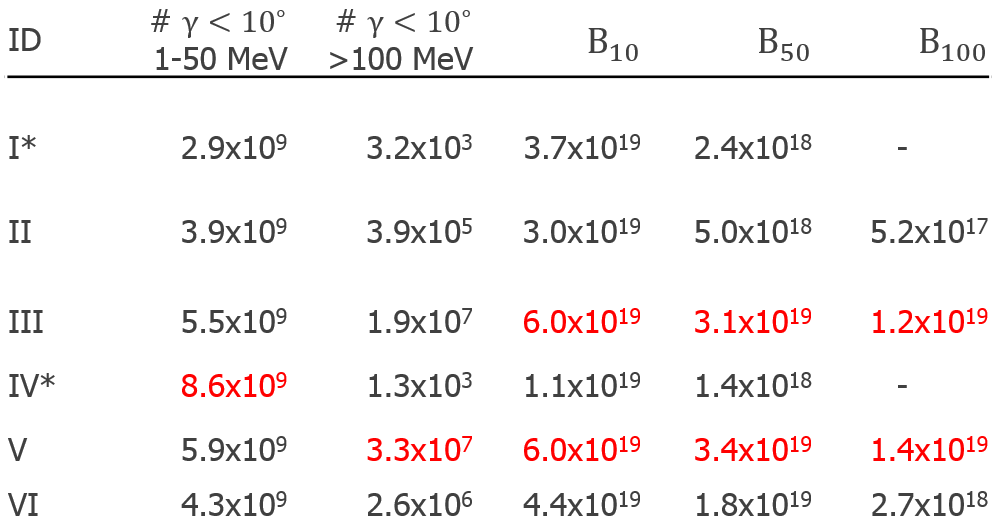}
\end{tabular}

* Radiation at or above 100MeV is too noisy to be reliable.
\caption{Collimation is evaluated for all 6 cases. In columns 2 and 3, number of photons inside a cone with a half angle of 10$^{\circ}$ is calculated within different energy ranges as indicated. Columns 4, 5 and 6 show the source brightness for Bremsstrahlung at energies of 10, 50 and 100MeV with 0.1$\%$ bandwidth. The unit is ${\rm s^{-1}mm^{-2}mrad^{-2}}$. The source diameter is estimated to be the FWHM of the laser focal spot. The pulse duration is taken as the pulse width of the corresponding electron bunch as described in the text. The brightness is averaged over the 10$^{\circ}$ half-angle cone.}
\end{table}

To further evaluate the Bremsstrahlung divergence, The number of photons as well as the average brightness within a small cone (10$^{\circ}$ half angle) are shown in Table \textcolor{blue}{II}. The brightness calculation uses an energy bandwidth of 0.1$\%$ and a source diameter of the laser focal spot FWHM and averaged over the 10$^{\circ}$ half cone. We cannot get  temporal information from the Monte Carlo simulation, so for the pulse width we estimate that the $\gamma$-ray and electron bunch pulse widths are the same, for each energy. For example, for $B_{100}$, the 100 MeV $\gamma$-rays could only have been produced by electrons whose kinetic energy is larger than 100 MeV so we take the bunch width of all electrons above that and use it for the corresponding $\gamma$-ray pulse width (the same is done for $B_{10}$ and $B_{50}$). This may be a slight underestimate of the $\gamma$-ray pulse width, but the error should be small. At each energy range, the electrons are all moving in excess of $99 \%$ of the speed of light and the emission angle of the radiation with respect to the electron direction is small. Hence, the radiation pulse will have similar temporal properties to the electron pulse.  The unit in Table \textcolor{blue}{II} is ${\rm s^{-1}mm^{-2}mrad^{-2}}$. 

Consistent with the results for electron generation, the low intensity tower target produces more total radiation but is the lowest in brightness for 10MeV and 50MeV due to a larger source size. The tower targets at high intensity have the highest brightness for all energies we consider. Finally, we note that the brightnesses do not rapidly decay for the tower targets at high intensity. For the best target (V), the brightness is 6.0$\times$10$^{19}$ ${\rm s^{-1}mm^{-2}mrad^{-2}}$ at 10MeV and 1.4$\times$10$^{19}$ ${\rm s^{-1}mm^{-2}mrad^{-2}}$ at 100MeV. It is likely that there will be lasers with the parameters we have considered that can operate at 10 Hz or greater in the near future. In that case, the average brightness of this source would be 2.3$\times 10^{7}$ ${\rm s^{-1}mm^{-2}mrad^{-2}}$ at 10 MeV and 2.1$\times 10^{6}$ ${\rm s^{-1}mm^{-2}mrad^{-2}}$ at 100 MeV.

\section{Conclusion}
\label{sec:conclusion}

We have proposed novel front surface target structures (towers and slabs) for the purpose of producing high energy, collimated electrons and $\gamma$-rays. \emph{Compared to regular flat targets with 1 $\mu$m pre-plasma, the yield of electrons at the high energy end can be improved by several orders of magnitude}, while the FWHM divergence angle of the most energetic hot electrons can be greatly reduced to $<$ 5$^{\circ}$. For electrons above 100 MeV, the average current can be as high as 70 kA. For high energy electrons (100 MeV), the brightness is on the order of ${\rm 10}^{23}$ ${\rm s^{-1}mm^{-2}mrad^{-2}}$ (0.1$\%$ bandwidth). Similarly, for high energy $\gamma$-rays (10 and 100 MeV), the brightness is as high as 6.0$\times$10$^{19}$ ${\rm s^{-1}mm^{-2}mrad^{-2}}$ (10 MeV) and  1.4$\times$10$^{19}$ ${\rm s^{-1}mm^{-2}mrad^{-2}}$ (100 MeV) (0.1$\%$ bandwidth) which is similar to other planned sources \textcolor{blue}{\cite{ELINP_Paper}}.

We have also performed a parameter scan to consider different target geometries and laser characteristics. As expected, the electron energy distribution is most sensitive to the laser intensity, however, the tower structured targets are shown to produce a similar distribution to a standard flat target at an order of magnitude lower intensity. Furthermore, the angular distribution is significantly narrowed leading to a significant narrowing of the electron beam. Naturally, these results are also observed in the energy and angular distributions of the $\gamma$-rays. We find that approximately $4-5\%$ of the laser energy can be converted into $\gamma$-rays and that these targets can be a bright source of high energy $\gamma$-rays.

\begin{acknowledgments}
This work is supported by the AFOSR Young Investigator Program (YIP) under contract FA9550-12-1-0341 as well as US Department of Energy contract DE-NA0001976. We would like to thank JT Morrison for useful discussions. Computational time was granted by the Ohio Supercomputer Center. 
\end{acknowledgments}

\bibliography{bibfile_structure}

\end{document}